\documentclass[12pt]{article}
\usepackage{latexsym}
\usepackage{amsmath,amssymb,amsfonts,amsthm}
\usepackage[english]{babel}
\newcommand{\be}{\begin{equation}}
\newcommand{\ee}{\end{equation}}
\newcommand{\bea}{\begin{eqnarray}}
\newcommand{\eea}{\end{eqnarray}}
\newcommand{\nn}{\nonumber \\}
\newcommand{\p}[1]{(\ref{#1})}
\newcommand{\lb}{\label}

\def\theequation{\arabic{section}.\arabic{equation}}
\begin{document}
\begin{titlepage}

\vfill

\begin{center}
\baselineskip=16pt {\Large\bf Bispinor Auxiliary Fields in Duality-Invariant}
\vspace{0.2cm}

{\Large\bf Electrodynamics Revisited: The $U(N)$ Case
}
\vskip 0.6cm {\large {\sl }} \vskip 10.mm {\bf E.A.
Ivanov, $\;$ B.M. Zupnik }
\vspace{1cm}

{\it Bogoliubov Laboratory of Theoretical Physics, JINR, \\
141980 Dubna, Moscow Region, Russia\\
}
\vspace{0.3cm}

{\tt eivanov@theor.jinr.ru},  $\;$ {\tt zupnik@theor.jinr.ru}
\end{center}
\vspace{1.6cm}

\par
\begin{center}
{\bf ABSTRACT}
\end{center}
\begin{quote}
We update and detail the  formulation of the duality-invariant
systems of $N$ interacting abelian gauge fields with $N$ auxiliary
bispinor fields added. In this setting, the self-duality amounts to $U(N)$ invariance of the
nonlinear interaction of the auxiliary fields. The $U(N)$ self-dual Lagrangians
arise after solving the nonlinear equations of motion for the auxiliary fields.
We also elaborate on a new extended version of the bispinor field
formulation involving some additional scalar auxiliary fields
and study $U(N)$ invariant interactions with derivatives of the auxiliary
bispinor fields. Such interactions generate higher-derivative $U(N)$ self-dual
theories.

\vspace{4.5cm}


\vfill \vfill \vfill \vfill \vfill
\end{quote}
\end{titlepage}

\setcounter{footnote}{0}

\setcounter{page}{1}

\section{Introduction}

The $U(N)$ duality property is inherent to nonlinear interactions of
$N$ abelian gauge
field strengths $F_{\alpha\beta}^k, \bar F_{\dot\alpha\dot\beta}^k$\,,
$(k =1, \ldots, N)$ \cite{GZ}-\cite{AFZ}.
The notorious examples of these duality invariant systems are
multi-field generalizations of the Born-Infeld (BI) theory. So far, the
construction of such generalized BI systems
was based on introducing some auxiliary matrix scalar fields $\chi$ with
bilinear algebraic relations between these fields and the scalar combinations
of the gauge field strengths
\cite{ABMZ}. The corresponding nonlinear Lagrangians as functions of
$F_{\alpha\beta}^k, \bar F_{\dot\alpha\dot\beta}^k$
arose after substituting the perturbative solution
for $\chi$.

The recent revival of interest in the duality-invariant systems was mainly triggered by the hypothesis
that the duality considerations could play the decisive role in checking the conjectured ultraviolet
finiteness of the maximally extended ${\cal N}=8, d=4$ supergravity (see, e.g., \cite{BHN,Ka,BN}). In theories
of this kind there simultaneously appear a few gauge fields, so it is just $U(N)$ duality that is
of relevance to this circle of problems.

The auxiliary {\it bispinor} field formulation of the $U(N)$ duality was
introduced in \cite{IZ1} as a natural generalization of the analogous
approach to the $U(1)$ duality-symmetric (or self-dual) systems \cite{IZ,IZ2}.
The $U(N)$ self-duality is equivalent to the manifest  $U(N)$ invariance of
the interaction Lagrangian for the auxiliary fields.

In this paper we further detail the bispinor field formulation and consider several
new examples of the $U(N)$ duality-invariant models. It is a continuation of
our recent paper \cite{IZ3}, where the bispinor
auxiliary field formulation of the $U(1)$ duality
was renewed and related to the latest developments in this area.

We start, in Section 2, by recalling the standard setting for the $U(N)$
self-dual theories in terms of $N$ Maxwell gauge field strengths
$F^k_{\alpha\beta}$ and then turn to the $(F,V)$
representation of these theories with $N$ bispinor auxiliary fields
$V^k_{\alpha\beta}$ added.
By construction, the $U(N)$ self-dual $(F,V)$ Lagrangian
satisfies the Gaillard-Zumino (GZ) representation with
an arbitrary invariant interaction ${\cal E}(V)$ \cite{IZ1}.
We study the general parametrization of the scalar $U(N)$ invariants
constructed out of the auxiliary fields. The  equations of motion for the
auxiliary fields are in one-to-one correspondence with
``the deformed twisted
self-duality constraints'' (for the $U(N)$ duality group) proposed in
\cite{BN}. We also introduce additional scalar matrix auxiliary fields $\mu$
and consider an alternative formalism involving two types of the
auxiliary fields. This $\mu$ representation simplifies solving
the auxiliary-field equations and constructing the
self-dual Lagrangians.

Section 3 is devoted to examples of the $U(N)$ self-dual theories, including $U(N)$ generalizations of the BI theory.
The first type of the $U(N)$ BI models
is the real form of the $U(N)\times U(N)$ generalization
considered in Ref. \cite{ABMZ}.
We translate this model into our $\mu$
representation with the special $U(N)$  invariant  auxiliary interaction.
 We also propose an alternative $U(N)$ generalization of the BI theory and construct the
perturbative Lagrangian for this model. Some other examples of the $U(N)$ self-dual models are as well considered.
One example corresponds to the simplest quartic interaction of the auxiliary
fields, and another one is constructed by analogy with the new exact $U(1)$
self-dual
Lagrangian given in \cite{IZ3}.

The $U(N)$ self-dual models with higher derivatives are studied in
Section 4. In the Appendix we rewrite our $(F^k,V^k)$ Lagrangian and
the self-duality equation in the  tensor formalism.

Throughout the paper we basically use the notations and abbreviations of
Ref. \cite{IZ3}.
\setcounter{equation}0
\section{Auxiliary fields for $U(N)$ duality}

\subsection{The standard setting}

Our starting point is the nonlinear Lagrangian with $N$ abelian
gauge field strengths
$F^i_{\alpha\beta},\,\bar{F}^i_{\dot\alpha\dot\beta}\,,\; (i = 1,
\ldots, N)$
\be
L(F^k,\bar{F}^l)=-\frac12[(F^kF^k)+(\bar{F}^k\bar{F}^k)]+L^{int}(F^k,\bar{F}^l)\,.
\ee
It is manifestly invariant under the real $O(N)$
transformation
\bea
&&\delta_{\xi} F^k_{\alpha\beta}=\xi^{kl}F^l_{\alpha\beta}\,,\quad
\delta_{\xi}\bar{F}^k_{\dot\alpha\dot\beta}=\xi^{kl}\bar{F}^k_{\dot\alpha\dot\beta}\,,\quad
\xi^{kl}=-\xi^{lk}\,.\lb{ON}
\eea
It is convenient to define $\frac12
N(N+1)$ complex scalar variables and to consider the Lagrangian as a
real function of these variables
\bea
&&\varphi^{kl}=\varphi^{lk}=(F^kF^l)\,,\quad
\bar\varphi^{kl}=(\bar{F}^k\bar{F}^l)\,,\\
&&L(F^k,\bar{F}^k)=L(\varphi^{kl},\bar\varphi^{kl})\,.
\eea

The nonlinear equations of motion
\bea
&&E^k_{\alpha\dot\alpha}=\partial_\alpha^{\dot\beta}
\bar{P}^k_{\dot\alpha\dot\beta}(F) -\partial^\beta_{\dot\alpha}
P^k_{\alpha\beta}(F)= 0 \lb{eomF}
\eea
involve the dual nonlinear field
strenghts\footnote{In the tensor notation one deals with the
self-dual field ${\tt G}^{+k}_{mn}\,$,
$$
P_{\alpha\beta}^{k}(F)=\frac18(\sigma^m\bar\sigma^n-
\sigma^n\bar\sigma^m)_{\alpha\beta}{\tt G}^{+k}_{mn}\,.
$$}
\bea P^k_{\alpha\beta}(F)=i\frac{\partial L}{\partial
F^{k\alpha\beta}} =2iF^l_{\alpha\beta}\,\frac{\partial L}{\partial
\varphi^{kl}}\,, \quad {\rm and\;\; c.c.}\,.
\eea
The gauge field strengths $F^{k}_{\alpha\beta}$, $\bar F^{k}_{\dot\alpha\dot\beta}$ obey the standard Bianchi identities
\bea
&&B^k_{\alpha\dot\alpha}=\partial_\alpha^{\dot\beta}
\bar{F}^k_{\dot\alpha\dot\beta} -\partial^\beta_{\dot\alpha}
F^k_{\alpha\beta}= 0\,. \lb{Bianchi}
\eea

The on-shell duality transformations are realized as
\be
\delta_{\eta}
F^k_{\alpha\beta}=\eta^{kl}P^l_{\alpha\beta}\,, \quad \delta_{\eta} P^k_{\alpha\beta}=-\eta^{kl}F^l_{\alpha\beta}\,, \lb{uN1}
\ee
where $\eta^{kl}=\eta^{lk}\,$ are $\frac12 N(N+1)$ real parameters. These transformations extend the $O(N)$ group \p{ON}
to the group $U(N)$ and so reside in the coset $U(N)/O(N)$. The equations of motion \p{eomF} together with the Bianchi identities \p{Bianchi}
are covariant under \p{uN1},
\be
\delta_{\eta} E^k_{\alpha\dot\beta} = -\eta^{kl}B^k_{\alpha\dot\beta}\,, \quad \delta_{\eta} B^k_{\alpha\dot\beta} = \eta^{kl}E^k_{\alpha\dot\beta}\,,
\ee
provided that the generalized consistency conditions hold:
\bea
(P^kP^l)+(F^kF^l)- {\rm c.c.} =0\,, \quad (F^kP^l)-(F^lP^k)-{\rm c.c.} = 0\,.
\lb{GZ}
\eea
These conditions are $U(N)$ covariant on their own right. The symmetric conditions
can be rewritten as the matrix differential equations for the Lagrangian
\be
\varphi-4L_\varphi\varphi L_\varphi=\bar\varphi-4L_{\bar\varphi}\bar\varphi
L_{\bar\varphi}\,.\lb{UNmatr}
\ee

\subsection{The $(F, V)$ representation}

The generalized auxiliary field representation of the $O(N)$
invariant Lagrangian \cite{IZ1} contains $N$ complex auxiliary
$O(N)$ vector fields $V^k_{\alpha\beta}\,$, $\bar V^k_{\dot\alpha\dot\beta}\,$
\bea
{\cal L}(F^k,V^k)&=&{\cal L}_2(F^k,V^k)+E[(V^kV^l),(\bar{V}^k\bar{V}^l)]\,,
\lb{LFVN}\\
{\cal L}_2(F^k,V^k) &=& \frac12[(F^kF^k)+(\bar{F}^k\bar{F}^k)]-2[(F^kV^k)
+(\bar{F}^k\bar{V}^k)] \nn
&& +\, (V^kV^k)+(\bar{V}^k\bar{V}^k)\,. \lb{L2FVN}
\eea
Here, $E$ is an $O(N)$ invariant real interaction depending on $\frac12 N(N+1)$
scalar complex variables
\bea
&&\nu^{kl}=(V^kV^l)\,, \quad \bar\nu^{kl}=(\bar V^k\bar V^l)\,.
\eea
The interaction $E(\nu^{kl},\bar\nu^{kl})$ is assumed to be regular at the origin,
so that it admits expansion in power series. The dynamical equation of motion following
from this Lagrangian has the form
\bea
\partial_\alpha^{\dot\beta}
\bar{P}^k_{\dot\alpha\dot\beta}(F,V)
-\partial^\beta_{\dot\alpha} P^k_{\alpha\beta}(F,V)= 0\,,\lb{EqFV}
\eea
where
\bea
P^k_{\alpha\beta}(F,V)=i(F^k-2V^k)_{\alpha\beta}\,.\lb{PFVdef}
\eea

The $U(N)/O(N)$ duality transformations are implemented as
\bea
&&\delta_{\eta}
F^k_{\alpha\beta}=\eta^{kl}P^l_{\alpha\beta}=i\eta^{kl}(F^l-2V^l)_{\alpha\beta}\,,\quad
\delta_{\eta} P^k_{\alpha\beta}=-\eta^{kl}F^l_{\alpha\beta}\,.\lb{eta1}
\eea
The corresponding $\eta^{kl}$ transformations of
the auxiliary fields follow from \p{eta1} and the definition \p{PFVdef}. The full $U(N)$
transformations of the auxiliary fields can be written as
\bea
&&\delta V^k_{\alpha\beta}=(\xi^{kl}-i\eta^{kl})V^l_{\alpha\beta}\,,
\quad
\delta\bar{V}^k_{\dot\alpha\dot\beta}=(\xi^{kl}+i\eta^{kl})
\bar{V}^l_{\dot\alpha\dot\beta}\,.
\eea
The $SU(N)$ subgroup is singled out by the condition
$\eta^{kk}=\mbox{Tr}\,\eta=0\,$.

The $U(N)$ transformations of the scalar variables $\nu^{kl}$ and
$\bar\nu^{kl}$
can be presented in the matrix form as
\bea
&&\delta\nu=[\xi,\nu]-i\{\eta,\nu\}\,,\quad
\delta\bar\nu=[\xi,\bar\nu]+i\{\eta,\bar\nu\}\,. \eea
Then, we define
the Hermitian matrix variables
\bea
&&a^{kl}=(\bar\nu\nu)^{kl}\,,\quad \bar{a}^{kl}=(\nu\bar\nu)^{kl} =
(\bar\nu\nu)^{lk}\,,\\
&& \delta a=i[\xi,a]+i[\eta,a]\,,\quad
\delta\bar{a}=i[\xi,\bar{a}]-i[\eta,\bar{a}]\,. \eea
We also define the matrix monomials $a^n$ with the following
properties
\bea
(a^n\bar\nu)^{kl}=(\bar\nu\bar{a}^n)^{kl}=(a^n\bar\nu)^{lk}\,,\quad
(\nu a^n)^{kl}=(\bar{a}^n\nu)^{kl}=(\nu a^n)^{lk}\,.\lb{anurel} \eea
{}From these monomials one can construct $N$ independent real $U(N)$
invariants $A_n\,$, $(n=1, 2,\ldots\,, N)\,$:
\bea
&&A_n=\frac1n\mbox{Tr}\,a^n,\qquad
dA_n=\mbox{Tr}\,(daa^{n-1})\,, \quad \frac{\partial A_n}{\partial
a^{kl}}=(a^{n-1})^{kl}\,.
\eea
An alternative choice of the $U(N)$ invariants is
connected with the spectrum
 $\lambda_1(A_1, \ldots\,, A_N)\,$, $\ldots, \lambda_N(A_1, \ldots\,, A_N)$
of the Hermitian matrix $a$ \cite{Ga,ZO}. This spectrum can be found
by solving the characteristic equation
\bea
A(a)=(a-\lambda_1)(a-\lambda_2)\cdots (a-\lambda_N)=0\,.
\eea

Like in the $U(1)$ case \cite{IZ,IZ2,IZ3}, the dynamical equations of motion
\p{EqFV}, together with
the Bianchi identities \p{Bianchi} and the algebraic equations of motion for
the auxiliary fields $V^k_{\alpha\beta}, \bar
V^k_{\dot\alpha\dot\beta}$ following from \p{LFVN}, are covariant under the
$U(N)$ duality transformations, provided that the interaction function
$E(\nu^{kl}, \bar\nu^{kl})$ in \p{LFVN} is $U(N)$ invariant \cite{IZ1},
\be
E(\nu^{kl},\bar\nu^{kl})~ \Rightarrow~ {\cal E}(A_1, \dots\,, A_N)\,, \lb{defcalE}
\ee
with ${\cal E}(A_1, \dots\,, A_N)$ being an analytic function.

The more convenient representation for the interaction Lagrangian
${\cal E}(A_n)$ is through the matrix function $E(a)$
\bea
&&{\cal E}=\mbox{Tr}\,E(a)\,.
\eea
The derivative matrix function $E_a$ is defined as follows
\bea
d{\cal E} = dA_n {\cal E}_n = \mbox{Tr}\,(da E_a)=
(da^{lk} E_a^{kl})\,,   \quad {\cal E}_n = \frac{\partial{\cal E}}{\partial A_n}\,, \lb{AB}
\eea
whence
\bea
(E_a)^{kl} =
{\cal E}_1\delta^{kl}+{\cal E}_2a^{kl}
+{\cal E}_3 a^{kj}a^{jl}+\ldots\,. \lb{AB2}
\eea
Using this representation and the
relations \p{anurel}, we define the holomorphic derivatives
\bea
&&{\cal E}^{kl}:= \frac{\partial{\cal E}}{\partial \nu^{kl}}=(E_a)^{kr}\bar\nu^{rl} =
{\cal E}_1\bar\nu^{kl}+{\cal E}_2a^{kr}\bar\nu^{rl}
+{\cal E}_3 a^{kj}a^{jr}\bar\nu^{rl}+\ldots,\\
&&\bar{\cal E}^{kl}:=\frac{\partial{\cal E}}{\partial \bar\nu^{kl}}=\nu^{kr}(E_a)^{rl}
={\cal E}_1\nu^{kl}+{\cal E}_2\nu^{kr}a^{rl} +{\cal E}_3
\nu^{kr}a^{rj}a^{jl}+\ldots\,.
\eea

The basic algebraic equations of the $U(N)$ duality-invariant models  are obtained by
varying, with respect to $V^k_{\alpha\beta}$
and $\bar{V}^k_{\dot\alpha\dot\beta}\,$, the Lagrangian \p{LFVN}, in which the general function $E$ is substituted
by the $U(N)$ invariant one ${\cal E}$ defined in \p{defcalE}:
\bea
(F^k-V^k)_{\alpha\beta}={\cal E}^{kl}V^l_{\alpha\beta}=(E_a)^{kr}
\bar\nu^{rl}V^l_{\alpha\beta}\,,\quad
(\bar{F}^k-\bar{V}^k)_{\dot\alpha\dot\beta}= \bar{\cal
E}^{kl}\bar{V}^l_{\dot\alpha\dot\beta}=\nu^{kr}(E_a)^{rl}
\bar{V}^l_{\dot\alpha\dot\beta}\,.\lb{EFVN}
\eea
Equations of motion \p{EFVN} are equivalent to the nonlinear twisted
self-duality constraints which were postulated in \cite{BN,CKR}.
The important corollaries of \p{EFVN} are the scalar matrix algebraic
equation
\bea
&&\varphi^{kl}=[\delta^{kr}+{\cal
E}^{kr}]\nu^{rs}[\delta^{sl}+{\cal E}^{sl}] \lb{varnukl}
\eea
and its conjugate.

By analogy with the $U(1)$ case \cite{IZ1,IZ3},   the general solution of the algebraic
equations \p{EFVN} can be
written in the following concise form:
\bea
&&V^k_{\alpha\beta}=F^l_{\alpha\beta}G^{kl}(\varphi,\bar\varphi)\,, \lb{VFG} \\
&&G^{kl}=[\delta^{kl}+{\cal E}^{kl}]^{-1}=\frac12\delta^{kl}
-\frac{\partial L}{\partial\varphi^{kl}}\,,\lb{GLkl} \\
&&P^k_{\alpha\beta}=2iF^l_{\alpha\beta}\,\frac{\partial
L}{\partial\varphi^{kl}} =iF^l_{\alpha\beta}[\delta^{kl}-2G^{kl}]\,.
\eea
Using \p{VFG}, \p{GLkl}, we can uniquely restore the Lagrangian
$L(\varphi^{kl},\bar\varphi^{kl})$ by its holomorphic derivatives:
\be
dL=d\varphi^{lk}\frac{\partial
L}{\partial\varphi^{kl}}+\mbox{c.c.}\,.
\ee
Note that Eqs. \p{EFVN}, \p{varnukl} are simplified under the
particular choices of ${\cal E}$, e.g., for ${\cal E} ={\cal
E}(A_1)$:
\bea
&&F^k_{\alpha\beta}=V^r_{\alpha\beta}[\delta^{kr}+{\cal E}_1
\bar\nu^{kr}]\,,\lb{EFVN2}\\
&&\varphi^{kl}
=\nu^{kl}+{\cal E}_1 [\bar\nu^{kr}\nu^{rl}
+\nu^{ks}\bar\nu^{sl}]+{\cal E}^2_1
\bar\nu^{kr}\nu^{rs}\bar\nu^{sl}\,.
\eea

The famous GZ representation of the $U(N)$ self-dual Lagrangians has the following form in the $(F, V)$ representation
\bea
{\cal L}(F^k,V^k)&=&\frac{i}2[\bar{P}^k(F,V)\bar{F}^k-P^k(F,V)F^k]
+[(V^kV^k)-(F^kV^k)]\nn
&&+\,[(\bar{V}^k\bar{V}^k)-(\bar{F}^k\bar{V}^k)]
+{\cal E}\,,\lb{GZFVN}
\eea
where $(V^kV^k)-(F^kV^k)$ is the complex
bilinear $U(N)$ invariant. Using Eqs. \p{EFVN}, we can also prove that
the $U(N)$ self-duality  conditions \p{GZ} in the $(F,V)$ representation are none other than
the conditions of $U(N)$ invariance of the auxiliary interaction ${\cal E}$ \cite{IZ1}
\bea
&& (P^kP^l)+(F^kF^l)- {\rm c.c.} =\left(V^l\frac{\partial{\cal E}}{\partial V^k}\right)+
\left(V^k\frac{\partial{\cal E}}{\partial V^l}\right)-{\rm c.c.} =0\,, \\
&& (F^kP^l)-(F^lP^k)-{\rm c.c.} = i\left(V^k\frac{\partial{\cal E}}{\partial V^l}\right)
- i\left(V^l\frac{\partial{\cal E}}{\partial V^k}\right) + {\rm c.c.}=0\,.
\eea

\subsection{The $\mu$ representation}

In the $U(1)$ case, there is a more convenient parametrization of the duality-invariant auxiliary interaction,
the ``$\mu$ representation'' \cite{IZ2,IZ3}. Making use of it essentially simplifies the road from the auxiliary-field equations
to the final nonlinear duality-invariant Lagrangian.

In the $U(N)$ case, the $\mu$ representation is set up  in
terms of the matrix variables
\bea
&&\mu^{kl}=\frac{\partial{\cal E}(a)}{\partial\nu^{kl}}= (E_a)^{kr}\bar\nu^{rl}\,,
\quad \bar\mu^{kl}=\nu^{kr}(E_a)^{rl}\,,\lb{mukl}\\
&& b^{kl}=\mu^{ks}\bar\mu^{sl}=(E_a\bar\nu\nu E_a)^{kl}=(aE^2_a)^{kl},\quad
b^{lk}=\bar{b}^{kl}=\bar\mu^{kr}\mu^{rl}\,, \lb{bkl}
\eea
where the
relations \p{anurel} were used. These newly defined matrix variables possess the
following transformation laws:
\bea
&&\delta
\mu=[\xi,\mu]+i\{\eta,\mu\},\quad\delta\bar\mu=[\xi,\bar\mu]-
i\{\eta,\bar\mu\},\quad \delta b=[\xi,b]+i[\eta,b]
\eea
and reveal the properties
\bea
(b^n\mu)^{kl}=(\mu\bar{b}^n)^{kl}=(b^n\mu)^{lk},\quad (\bar\mu
b^n)^{kl}=(\bar{b}^n\bar\mu)^{kl}= (\bar\mu b^n)^{lk}\,.\lb{bmurel}
\eea
{}From them one can construct $N$ independent $U(N)$ invariants
\be
B_n=\frac1n\mbox{Tr}\,b^n\,,\lb{Binv}
\ee
which are going to be the arguments of the $\mu$ representation analog of
the invariant function ${\cal E}(A_n)\,$.

The connection with the basic objects of the original $\nu$ representation
is established through the Legendre transformation
\bea
&&{\cal I}(B_n) = \mbox{Tr}\,I(b) :={\cal
E}-\nu^{kl}\mu^{kl}-\bar\nu^{kl}\bar\mu^{kl}=\mbox{Tr}\,[E-2aE_a]\,,\\
&&\nu^{kl}=-\frac{\partial{\cal I}}{\partial\mu^{kl}}=-(\bar\mu I_b)^{kl}=
-(\bar{I}_{\bar{b}}\bar\mu)^{kl}\,,\\
&&d{\cal I}=\mbox{Tr}\,(db I_b)=\mbox{Tr}\,(d\bar{b}
\bar{I}_{\bar{b}})\,,
\eea
where ${\cal I}(B_n)$ is a real analytic invariant function which is the interaction in the
$\mu$ representation. We introduced the covariant matrix functions $I(b)$ and $I_b$
which are  representable
as formal series over the powers of the matrix $b$,
with the coefficients being functions of the invariants $B_n$, e.g.,
\bea
I(b)=\sum_k \frac1{k!}I^{(k)}(B_n)\,b^k\,,\quad \delta I(b)=[(\xi+i\eta),I(b)]\,.
\eea
They are related to the matrix functions $E(a)$ and $E_a$ by the covariant
 matrix equations
\bea
I(b)=E(a)-2aE_a\,,\quad E(a)=I(b)-2bI_b\,,\quad E_a=-I^{-1}_b\,,\quad a=bI^2_b,\;\;
b=aE^2_a\,.
\eea
Eqs. \p{varnukl} can be rewritten in the $\mu$ representation as
\bea
\varphi^{kl}&=&[\delta^{kr}+{\cal E}_{kr}]\nu^{rs}[\delta^{sl}+{\cal E}_{sl}]
=-(\delta^{kr}+\mu^{kr})\frac{\partial{\cal I}}{\partial\mu^{rs}}
(\delta^{sl}+\mu^{sl})\nn
&=& -\,[({\bf 1}+\mu)\bar\mu I_b({\bf 1}+\mu)]^{kl}\,,
\eea
or
\bea
\varphi = -\,[({\bf 1}+\mu)\bar\mu I_b({\bf 1}+\mu)]\,, \quad \bar\varphi=-({\bf 1}+\bar\mu) I_b\mu({\bf 1}+\bar\mu)\,, \lb{varmukl}
\eea
where the relations \p{bmurel} are used and ${\bf 1}$ denotes the unit matrix.

In the particular representation, $I_b$ can be chosen as  the matrix
power series expansion with
the numerical coefficients $i_k$
\bea
&&
(I_b)^{kl}=-2\delta^{kl}+i_2b^{kl}+\frac12i_3(b^2)^{kl}+\ldots\,.
\eea
Then we can write the following recursive complex matrix equation for $\mu$ :
\bea
\mu &=& \frac12\bar\varphi-\mu\bar\mu-\bar\mu\mu-\bar\mu\mu\bar\mu+\frac12i_2\mu\bar\mu\mu
+\frac12i_2\bar\mu\mu\bar\mu\mu\nn
&&+\,\frac12i_2\mu\bar\mu\mu\bar\mu+
\frac12i_2\bar\mu\mu\bar\mu\mu\bar\mu+\frac14i_3\mu\bar\mu\mu\bar\mu\mu+O(\mu^6)\,.\lb{Irecur}
\eea
Solving it, e.g.,  for $\mu$ as  $\mu = \mu^{kl}(\varphi,\bar\varphi)\,$, we
can reconstruct the holomorphic derivatives of the Lagrangian from
Eqs. \p{GLkl} and \p{mukl},
\bea
\left(\frac{\bf 1}{{\bf 1}+\mu}\right)^{kl}=\frac12\delta^{kl}-
\frac{\partial L}{\partial\varphi^{kl}}\,,\lb{Lmukl}
\eea
and finally restore the  nonlinear perturbative Lagrangian in the $F$-representation
\bea
L&=&\frac12\mbox{Tr}\left[-(\varphi+\bar\varphi)+\varphi\bar\varphi
-\frac12\varphi\bar\varphi^2-\frac12\varphi^2\bar\varphi\right]\nn
&&+\,\frac18\mbox{Tr}\left[\varphi^3\bar\varphi
+\varphi\bar\varphi^3
+(2+\frac14i_2)\varphi\bar\varphi\varphi\bar\varphi
+2\varphi^2\bar\varphi^2\right]+O(\varphi^5)\,,\lb{LIpert}
\eea
where the  single-trace matrix  terms of higher orders are omitted.

Like in the $U(1)$ case \cite{IZ3}, one can define a combined $(F,V,\mu)$ off-shell representation for the $U(N)$ self-dual
Lagrangians, treating $\mu^{ik}, \bar\mu^{ik}$ as independent auxiliary fields
\bea
L(V^k,F^k,\mu^{kl})&=& \frac12[(F^kF^k)+(\bar{F}^k\bar{F}^k)]-
2\,[(V^k\cdot F^k)+(\bar{V}^k\cdot\bar{F}^k)]\nn
&& +\,(V^kV^l)(\delta^{kl}+\mu^{kl})+ \bar{V}^k\bar{V}^l
(\delta^{kl}+\bar\mu^{kl})+{\cal I}(B_n)\,,
\eea
where $B_n$ are the invariants \p{Binv}. Eliminating the $V^k$ variables from this
Lagrangian,
\be
V^k_{\alpha\beta}=\left[({\bf 1}+\mu)^{-1}\right]^{kl}F^l_{\alpha\beta}\,, \quad {\rm and \; c.c.}\,,
\ee
we arrive at the $(F,\mu)$ representation of the Lagrangian:
\bea
&&\tilde{L}(F^k,\mu^{kl})=\frac12
(F^{k}F^l)[(\mu-{\bf 1})({\bf 1}+\mu)^{-1}]^{kl} +\mbox{c.c.}+{\cal I}(B_n)\,.
\lb{tildeL}
 \eea
Varying this Lagrangian with respect to $\mu^{kl}$  we obtain the
matrix auxiliary equation which is equivalent to Eq. \p{varmukl}.

In the specific examples we can exploit the similarity between the $U(1)$ interaction function
$I(b)$ \cite{IZ3} and the matrix function $I(b)$ of the $U(N)$ case, although
solving the matrix equations is the much more difficult task. For the
simple  particular $U(N)$ interaction presented by a one-argument
function ${\cal E}(A_1)\,,\; A_1=a^{kk}=\nu^{kl}\bar\nu^{lk}\,$,
 we  find, e.g.,
\bea
&&\mu^{kl}={\cal E}_1\bar\nu^{kl}\,,\quad \bar\mu^{kl}={\cal
E}_1\nu^{kl}\,,\quad
\nu^{kl}=-{\cal I}_1\bar\mu^{kl}\,,\\
&&b^{kl}={\cal E}_1^2a^{kl},\quad B_1=b^{kk}={\cal
E}_1^2A_1,\quad{\cal E}_1=-{\cal I}^{-1}_1\,.
\eea
The corresponding
interaction function in the $\mu$-representation involves only the
trace $B_1\,$,
\bea
{\cal I}(B_1)={\cal E}(A_1)-2A_1{\cal
E}_1\,,\qquad \nu^{kl}=-{\cal I}_1\bar\mu^{kl},\quad a^{kl}={\cal
I}^2_1b^{kl}\,, \quad{\cal E}_1=-{\cal I}^{-1}_1\, .
\eea
The equation \p{varmukl} has the following form in this case:
\bea
\varphi=-{\cal I}_1({\bf 1}+\mu)\bar\mu({\bf 1}+\mu)\,.\lb{B1recur}
\eea
We consider the representation
\be {\cal
I}(B_1)=-2B_1+\frac12j_2B^2_1+\frac16j_3B^3_1+\ldots\,,\quad{\cal
I}_1=-2+j_2B^2_1+\frac12j_3B^2_1+\ldots \lb{B1pert}
\ee
($j_2, j_3,\ldots$ are some constants) and the corresponding recursion
relations for $\mu(\varphi,\bar\varphi)$. The 4-th order term in the
corresponding self-dual Lagrangian
\bea
&&L^{(4)}(\varphi,\bar\varphi)=\frac18\mbox{Tr}\left[\varphi^3\bar\varphi
+\varphi\bar\varphi^3 +2\varphi\bar\varphi\varphi\bar\varphi
+2\varphi^2\bar\varphi^2\right]+\frac1{32}j_2[\mbox{Tr}(\varphi\bar\varphi)]^2\,.\lb{LIpert1}
\eea
contains the double-trace term. The subsequent recursions give terms with
several traces.
\setcounter{equation}0

\section{Examples of the $U(N)$ self-dual models}
Here we present some examples of $U(N)$  self-dual models with actions involving no higher derivatives. Basically, these are generalizations of the
$U(1)$ examples considered in \cite{IZ3}. Similarly to the $U(1)$ case, the corresponding interactions written in terms of the auxiliary
variables can be chosen in a closed form, while the equivalent on-shell expressions, with the auxiliary variables being eliminated in terms
of the Maxwell field strengths, can be given only as infinite series in powers of the field strengths. An important difference from the $U(1)$ case
is that there exist several inequivalent $U(N)$ duality-invariant models which are reduced to the same $U(1)$ model in the one field-strength limit.
For instance, there are few $U(N)$ duality-invariant extensions of the standard BI theory. As distinct from the latter, the Lagrangians of such
generalized BI theories seem not to admit a closed representation in terms of the Maxwell field strengths (even for the simplest non-trivial $U(2)$ case).

\subsection{$U(N)$ generalizations of the Born-Infeld model}
The $U(N)\times U(N)$   generalization of the BI theory
proposed in \cite{ABMZ} deals with the  Hermitian scalar matrix fields
 \bea
 \hat\alpha^{lk}=\frac14({\tt F}^k\bar{\tt F}^l),\quad \hat\beta^{kl}=
 \frac14(\tilde{\tt F}^k
 \bar{\tt F}^l)
 \eea
constructed out of $N$ {\it complex} field-strengths
 ${\tt F}^k_{mn}$ and their conjugates $\bar{\tt F}^k_{mn}$ (with
 $\tilde{\tt F}^k_{mn} = \frac12 \varepsilon_{mnst}{\tt F}^{k\,st}$).
The basic complex scalar auxiliary field $\chi^{kl}\neq\chi^{lk}$ of this model
satisfies the  matrix equation
\bea
\chi^{kl}+\frac12\chi^{kr}\bar\chi^{lr}=\hat\alpha^{lk}+i\hat\beta^{kl}=
\hat\varphi^{kl}\,,
\eea
with the solution representable as the matrix power series.

The $U(N)$ generalization of the BI theory we are interested in corresponds to imposing
the reality condition ${\tt F}^k=\bar{\tt F}^k$ and using
the symmetric  matrices
\bea
&&\hat\alpha^{kl}\rightarrow t^{kl}=\frac14\eta^{mr}\eta^{ns}{\tt F}_{mn}^k
{\tt F}_{rs}^l\,,\quad
\hat\beta^{kl}\rightarrow z^{kl}=\frac18\varepsilon^{mnrs}{\tt F}_{mn}^k
{\tt F}_{rs}^l\,,\nn
&&\hat\varphi^{kl}~\rightarrow ~\varphi^{kl}=(F^kF^l)\,.
\eea
In this notation, we consider the following nonlinear Lagrangian of $N$ abelian
gauge field strengths $F_{\alpha\beta}^k$
\bea
L_{ABMZ}(\varphi,\bar\varphi)&=& \mbox{Tr}\left[-\frac12(\varphi+\bar\varphi)
+\frac12\varphi\bar\varphi-\frac14(\varphi^2\bar\varphi
+\varphi\bar\varphi^2)\right]\nn
&&+\,\frac1{8}\mbox{Tr}\left[\varphi^3\bar\varphi
+2\varphi^2\bar\varphi^2+\varphi\bar\varphi\varphi\bar\varphi
+\varphi\bar\varphi^3\right]+O(\varphi^5)\,.\lb{LABMZ}
\eea
The $U(N)$ duality condition \p{UNmatr} can be directly proved for
 this Lagrangian.

Our interpretation of this model makes use of the following exact invariant interaction in the
matrix $\mu$ representation
\p{tildeL}:
\bea
&&{\cal I}=\mbox{Tr}I(b)\,,\quad I(b)=\frac{2b}{(b-{\bf 1})}\,,\quad
I_b=-\frac2{(b-{\bf 1})^2}\,.
\eea
We can calculate $L(\varphi,\bar\varphi)$ as the power series, based on the expansion
\bea
I(b)=-2b-2b^2-2b^3-2b^4-\ldots\,.
\eea
For the proper choice of the numerical coefficients in \p{LIpert},  $i_2=-4,\ldots\,,$  we can reproduce the
 Lagrangian \p{LABMZ}.

By analogy with the $U(1)$ case, we can come back to the original $(F,V)$ formulation, defining the matrix variable $a$ by the
relation
\be
a = \frac{4b}{({\bf 1}-b)^4}\,.
\ee
The auxiliary interaction has the single-trace form ${\cal E}(a)=\mbox{Tr}E(a)$.
The matrix relations between various quantities in the $(F,V)$ and $\mu$ representations are similar to those valid
in the $U(1)$ duality case for the  BI theory  \cite{IZ2,IZ3}
\bea
&& E(a) = \frac{2b(a)[{\bf 1}+b(a)]}{[{\bf 1}-b(a)]^2} = 2[2t^2(a) + 3t(a)
+{\bf 1}]\,, \lb{Ebi}\\
&&
t^4 + t^3 -\frac14a =0, \quad t=\frac1{b-{\bf 1}}\,,\\
&&2 E_a=[{\bf 1}-a E_a^2]^2,\quad E_a=\frac12[b(a)-{\bf 1}]^2\,.\lb{Eder}
\eea
Solving these equations, one can find closed expressions for both $t(a)$ and
the single-trace interaction ${\cal E}(a)$. They look rather bulky and so it is not too illuminating to present them here.
Up to the 3-d order in $a$ :
\be
{\cal E} =\mbox{Tr}\left[ \frac12\,a - \frac18\,a^2 +
\frac{3}{32}\,a^3 + O(a^4)\right]. \lb{3BI}
\ee

An alternative $U(N)$ generalization of the BI Lagrangian proceeds from the $\mu$
 representation with the invariant auxiliary interaction of the simple form
\bea
&&{\cal I}(B_1)=\frac{2B_1}{B_1-1}\,,\quad B_1=\mbox{Tr}\mu\bar\mu\,.
\lb{LT1off}
 \eea
This interaction corresponds to the choice $j_2=-4,\ldots$ in \p{B1pert}.
We can obtain the self-dual
nonlinear Lagrangian in the $F$ representation, using the recursion equation
\p{B1recur}. The $(F,V)$ representation of the same model deals with the invariant
interaction which is a function of the single $U(N)$ invariant variable $A_1=(V^kV^l)(\bar{V}^l\bar{V}^k)$:
\be
{\cal E}(A_1) = \frac12\,A_1 - \frac18\,A_1^2 +
\frac{3}{32}\,A_1^3 + O(A_1^4)\,. \lb{3BIal1}
\ee

\subsection{Other examples of $U(N)$ self-dual theories}

The simplest quartic $U(N)$ invariant interaction
\bea
&&{\cal E}_{SI}=\frac12(V^kV^l)(\bar{V}^k\bar{V}^l)=\frac12\mbox{Tr}\,a
=\frac12A_1
\eea
produces the self-dual model, which is $U(N)$ generalization of the
``simplest interaction $U(1)$ self-dual model''
of refs. \cite{IZ2,IZ3} (it was rediscovered in \cite{BN,CKR}).
In this case, the basic polynomial auxiliary equation
\bea
&&F^k_{\alpha\beta}=[\delta^{kl}+\frac12(\bar{V}^k\bar{V}^l)]V^l_{\alpha\beta}
\eea
has the perturbative solution for the function $G_{kl}=\frac12\delta_{kl}
-\partial L_{SI}/\partial\varphi^{kl}$ \p{GLkl}. The corresponding power-series
Lagrangian reads
\bea
L_{SI}&=&\mbox{Tr}[-\frac12(\varphi+\bar\varphi)+\frac12\varphi\bar\varphi
-\frac14(\varphi^2\bar\varphi+\varphi\bar\varphi^2) \nn
&& +\,\frac18(\varphi^3\bar\varphi+2\varphi^2\bar\varphi^2
+2\varphi\bar\varphi\varphi\bar\varphi
+\varphi\bar\varphi^3)]+\ldots\,.
\eea
We can also consider  the $\mu$ representation for this model
\bea
&&{\cal I}(B_1)=-2B_1\,.
\eea

Some other $U(1)$ examples considered in \cite{IZ2,IZ3} also admit extensions
to the $U(N)$ duality case. For instance, we can choose the interaction
\bea
{\cal I}(B_1)=2\ln(1-B_1)=-2(B_1+\frac12B^2_1+\frac13B^3_1+\ldots)\,,\quad
{\cal I}_1=\frac2{B_1-1}
\eea
and construct the self-dual Lagrangian in the $F$-representation, using the perturbative solution of
Eq. \p{B1recur} specialized to this case.

\setcounter{equation}0
\section{Interactions with higher derivatives}
The generalized $U(N)$ self-dual Lagrangians with higher derivatives \cite{CKO} in the formulation through bispinor auxiliary fields
are constructed in the close analogy with the $U(1)$ case \cite{IZ3}. They involve the same bilinear term \p{L2FVN}  and
the $U(N)$ invariant  interaction ${\cal E}^K_{der}$ containing derivatives of the auxiliary fields
\bea
{\cal L}_{der}(F^k,V^k)={\cal L}_2(F^k,V^k)+{\cal E}^K(V,\partial V,
\partial^2 V,\ldots\partial^K V)\,,
\eea
where $K$ denotes the maximal total degree of derivatives. Terms with derivatives in
${\cal E}^K$ contain the  coupling constant $c$ of
dimension $-2$ and additional dimensionless coupling constants.
This generalized Lagrangian admits
the same GZ-representation \p{GZFVN}.

The $U(N)$-covariant local equations of motion for the auxiliary fields in this
case contain the Lagrangian derivative of
the invariant auxiliary interaction
\bea
(V^k-F^k)_{\alpha\beta}+\frac12\frac{\Delta{\cal E}^K}
{\Delta V^{k\alpha\beta}}=0\,,\lb{dertwisd}
\eea
where we consider the Lagrange derivative
\bea
&&\frac{\Delta{\cal E}^K} {\Delta V^{k}}=\frac{\partial{\cal E}^K} {\partial V^{k}}
-\partial_m\frac{\partial{\cal E}^K} {\partial \partial_mV^{k}}+\partial_m\partial_n
\frac{\partial{\cal E}^K} {\partial \partial_m\partial_nV^{k}}+\ldots.\nonumber
\eea
The equivalent twisted self-duality relations for higher-derivative
theories were postulated in \cite{BN,CKO}.

The consistency condition
\bea
&&(P^kP^l)+(F^kF^l)=2(F^kV^l)+2(F^lV^k)-4(V^kV^l)=\left(V^l
\frac{\Delta{\cal E}^K}{\Delta V^k}\right)+
\left(V^k\frac{\Delta{\cal E}^K} {\Delta V^l}\right)\nn
&&=\left[\left(V^k\frac{\partial{\cal E}^K}{\partial V^l}\right)
+\left(\partial_mV^k\frac{\partial{\cal E}^K}{\partial(\partial_m V^l)}\right)
+\left(\partial_m\partial_nV^k\frac{\partial{\cal E}^K}{\partial(\partial_m \partial_nV^l)}\right)+\ldots\right]
+[k\leftrightarrow l]\nn
&&
+\mbox{total derivatives}
\eea
is evidently valid for the  derivative interaction ${\cal E}^K\,$. Then the $\eta^{kl}$ invariance of the interaction
${\cal E}^K$ is equivalent to the following
integral form of the
 self-duality condition
\bea
\eta^{kl}\int
d^4x[(P^kP^l)+(F^kF^l)-(\bar{P}^k\bar{P}^l)-(\bar{F}^k\bar{F}^l)]=
\int d^4x[\delta_\eta{\cal E}^K+\mbox{derivatives}]=0\,.
\eea
The $O(N)$ $\xi$-invariance of ${\cal L}_{der}(F^k,V^k)$ is manifest, as in the case without derivatives.

Solving Eq. \p{dertwisd}, we obtain the perturbative solution
$V^k_{\alpha\beta}(F^k)$, which involves both the field strengths and their derivatives.

The simplest bilinear invariant with two derivatives
\bea
&&{\cal
E}(V, \partial V) =
ca_1\partial_\beta^{\dot\beta}V^{k\alpha\beta}\partial^{\dot\xi}_{\alpha}
\bar{V}^k_{\dot\beta\dot\xi}\,
\eea
makes the fields $V$ and $\bar{V}$ propagating.

An example of the nonlinear interaction with two derivatives (still with the standard bilinear terms)   corresponds
to the choice
\bea
&&{\cal E}^2=b_1c\partial^m(V^kV^l)\partial_m(\bar{V}^k\bar{V}^l)\,.
 \eea
The basic auxiliary equation in this case reads
\bea
&&F^k_{\alpha\beta}=V^l_{\alpha\beta}[\delta^{kl}-cb_1\Box
(\bar{V}^k\bar{V}^l)]\,,
\eea
and it can be recursively solved for $V^l_{\alpha\beta}$ in terms of $F^k_{\alpha\beta}$ and its derivatives.
The corresponding Lagrangian in the $F$-representation is given by the formal series
\bea
L=-\frac12(\varphi^{kk}+\bar\varphi^{kk})
+cb_1\partial^m\varphi^{kl}\partial_m\bar\varphi^{kl}
-c^2b^2_1(\varphi^{kl}\Box\bar\varphi^{lr}\Box\bar\varphi^{rk}
+ \bar\varphi^{kl}\Box\varphi^{lr}\Box\varphi^{rk})+O(c^3).
\eea
The number of derivatives increases with each recursion.

An example of the $U(N)$ invariant auxiliary interaction involving four
derivatives is
\bea
{\cal E}^4=gc^2(\partial^mV^k\partial^nV^l)(\partial_m\bar{V}^k
\partial_n\bar{V}^l)\,,
\eea
where $g$ is a dimensionless coupling constant.
The corresponding auxiliary field equation reads
\bea
&&F^k_{\alpha\beta}=V^k_{\alpha\beta}-2gc^2\partial^m\left[
\partial^nV^l_{\alpha\beta}(\partial_m\bar{V}^k\partial_n\bar{V}^l)\right].
\eea
It is not difficult to recursively solve this equation too and to construct
the corresponding self-dual Lagrangian with higher derivatives.

All these examples are $U(N)$ generalizations of the $U(1)$ self-dual Lagrangians with
higher derivatives presented in \cite{IZ3}. Like in the case without higher derivatives,
the set of inequivalent $U(N)$ duality-invariant models of this kind is much richer compared to their
$U(1)$ prototypes due to the proliferation of the Maxwell field strengths and the associated  auxiliary tensorial fields.
\setcounter{equation}{0}

\section{Conclusions}

In this paper, we further elaborated on the formalism with the bispinor (tensor)
auxiliary fields for the $U(N)$ self-dual abelian gauge theories initiated in
\cite{IZ1}. The general Lagrangian
of the $U(N)$ self-dual model is parametrized by the invariant
interaction of the auxiliary fields. The $U(N)$ covariant local twisted
self-duality condition arises  in this formulation as the equation of motion for the bispinor auxiliary fields.
As compared to \cite{IZ1},  we presented an alternative formulation of the
$U(N)$ self-dual theories which makes use of the matrix scalar auxiliary fields in parallel with the bispinor ones, discussed a few
new examples and showed how to generate  $U(N)$ self-dual theories with higher derivatives in the considered setting.

In a recent paper \cite{ILZ} we gave basic elements of ${\cal N}=1$ supersymmetric
generalization of the $U(N)$ self-dual bosonic actions, using the auxiliary chiral superfields.
The ${\cal N}=1$ supersymmetrization of the bispinor auxiliary field formalism for the $U(1)$ case, through enhancing this field to an
auxiliary chiral ${\cal N}=1$ superfield, was earlier  accomplished by Kuzenko \cite{Ku} within a more general framework
of the superfield ${\cal N}=1$ supergravity \footnote{In \cite{Ku}, ${\cal N}=2$ supersymmetrization was also considered.}.
The $U(N)$ self-duality formulations detailed here seem to admit rather straightforward supersymmetric extensions
(with both rigid and local supersymmetries) along the lines of these works.

We also note that the bispinor auxiliary field formulation can be set up as
well for self-dual abelian gauge theories in the $d=4$ Euclidean
space \cite{Eucl} and the space with the signature $(2, 2)$. The
complex fields $V_{\alpha\beta}^i$ and $\bar
V_{\dot\alpha\dot\beta}^i$ are substituted by two sets of real
independent fields transforming as $SO_L(3)$ and $SO_R(3)$ vectors
in the Euclidean case, or as $SO_L(1,2)$ and $SO_R(1,2)$ vectors for
the signature $(2, 2)$. The relevant duality group is non-compact,
and it is the general linear group $GL(N)$ (it is reduced to
dilatations $L(1)\sim SO(1,1)$ in the $N=1$ case).

Finally, it is worthwhile to mention that the notion of self-duality can be defined for theories with $p$-form gauge fields,
not only for $p=1$, and in diverse dimensions, not only for $d=4$ (see, e.g., \cite{KT} and references therein). It would be tempting to
introduce the appropriate tensorial auxiliary fields for this web of generalized self-dualities and to see how they could help in constructing
the relevant actions and understanding the interrelations between various types of such dualities
\footnote{It was shown, e.g., in \cite{Berm} that the $d=4$ BI theory can be obtained by dimensional reduction
from a self-dual theory in $d=6$.}.

\section*{Acknowledgements}
We acknowledge a partial support from the RFBR grants Nr.12-02-00517,
Nr.11-02-90445, the grant DFG LE 838/12-1 and a grant of the Heisenberg-Landau
program. E.I. thanks the Organizers of the Workshop ``Higher Spins, Strings and Duality'' (Galileo Galilei Institute for Theoretical Physics,
Florence, March 18 - May 10, 2013) for the kind hospitality at the final stage of this work, and the participants for useful discussions.

\setcounter{equation}0
\renewcommand{\theequation}{A.\arabic{equation}}
\section*{A. Spinor and tensor notations in self-dual theories}

Our bispinor $U(1)$ formalism translated to the tensor notation was considered
in \cite{IZ3}. Here we present the basic formulas of the tensor reformulation of this approach
for the case of $U(N)$ self-dual theories.

The vectors in the spinor and tensor notations are related as
\be
A^k_{\alpha\dot\beta}=(\sigma^m)_{\alpha\dot\beta}A^k_m\,,\lb{A1}
\ee
where $k=1,2,\ldots N$.
The same correspondence for $N$ abelian field strengths
is given by the relations
\bea
&&F_\alpha^{k\,\beta}(A)=
\frac14(\sigma^m\bar\sigma^n)_\alpha^\beta {\tt F}^k_{mn}
=\,\frac18(\sigma^m\bar\sigma^n-\sigma^n\bar\sigma^m)_\alpha^\beta
{\tt F}^{+k}_{mn}\,,\lb{A2}\\
&&\bar{F}_{\dot\alpha}^{k\,\dot\beta} =
\frac14(\bar\sigma^n)^{\dot\beta\beta}(\sigma^m)_{\beta\dot\alpha}{\tt F}^k_{mn}
=\,-\frac18(\bar\sigma^m\sigma^n
-\bar\sigma^n\sigma^m)^{\dot\beta}_{\dot\alpha}{\tt F}^{-k}_{mn}\,,\lb{A3}
\eea
where
\bea
&&{\tt F}^k_{mn}=\partial_mA^k_n-\partial_nA^k_m\,,\quad
\tilde{\tt F}^k_{mn}=\frac12\varepsilon_{mnrs}
{\tt F}^{krs}\,,\quad {\tt F}^{+k}_{mn}=\frac12{\tt F}^k_{mn}+\frac{i}2
\tilde{\tt F}^k_{mn}\,, \\
&&\widetilde{{\tt F}^{+k}}_{mn}=-i{\tt F}^{+k}_{mn}\,,\quad
{\tt F}^{-k}_{mn}=\frac12{\tt F}^k_{mn}-\frac{i}2\tilde{\tt F}^k_{mn}\,,\quad
\widetilde{{\tt F}^{-k}}_{mn}=i{\tt F}^{-k}_{mn}\,.\lb{A4}
\eea
Thus, $F^k_{\alpha\beta}$ is the equivalent bispinor notation for the
self-dual tensor field
${\tt F}^{+k}_{mn}\,$, and $\bar{F}^k_{\dot\alpha\dot\beta}$ amounts to
the anti-self-dual tensor ${\tt F}^{-k}_{mn}\,$.

The scalar variables in the spinor formalism are related to the
analogous variables in the tensor notation as
\bea
&&\varphi^{kl}=
F^{k\alpha\beta} F^l_{\alpha\beta}=t^{kl}+iz^{kl}=
\frac12({\tt F}^{+k}{\tt F}^{+l})\,,\nn
&&\bar\varphi^{kl}=\bar{F}^{k\dot\alpha\dot\beta}\bar{F}^l_{\dot\alpha\dot\beta}
=t^{kl}-iz^{kl}=\frac12({\tt F}^{-k}{\tt F}^{-l})\,.\lb{A5}
\eea

The bispinor and tensor representations of the dual field strengths
appearing in the nonlinear equations of motion are related as
\bea
&&P_\alpha^{k\beta}(F)=\frac18(\sigma^m\bar\sigma^n-
\sigma^n\bar\sigma^m)_\alpha^\beta {\tt G}^{+k}_{mn}\,,\quad
\bar{P}_{\dot\alpha}^{k\dot\beta}=-\frac18(\bar\sigma^m\sigma^n-
\bar\sigma^n\sigma^m)_{\dot\alpha}^{\dot\beta}{\tt G}^{-k}_{mn}\,,\lb{A7}\\
&&{\tt G}^{\pm k}_{mn}=\frac12{\tt G}^k_{mn}\pm \frac{i}2\tilde{\tt G}^k_{mn}\,,\quad
\tilde{\tt G}^k_{mn}= 2\,\frac{\Delta L} {\Delta {\tt F}^{kmn}}\,,
\eea
where we employed the Lagrange derivative.

The similar relations are valid for the auxiliary fields
\bea
V_\alpha^{k\beta}=\frac18(\sigma^m\bar\sigma^n-
\sigma^n\bar\sigma^m)_\alpha^\beta {\tt V}^{+k}_{mn}\,,\qquad
\bar{V}_{\dot\alpha}^{k\dot\beta}=-\frac18(\bar\sigma^m\sigma^n-
\bar\sigma^n\sigma^m)_{\dot\alpha}^{\dot\beta} {\tt V}^{-k}_{mn}\,.\lb{A9}
\eea
The scalar
variable $\nu$ can be expressed as
\bea
\nu^{kl}=V^{k\alpha\beta}V^l_{\alpha\beta}=\frac12({\tt V}^{+k}{\tt V}^{+l})\,,
\quad
\bar\nu^{kl}=\bar{V}^{k\dot\alpha\dot\beta}\bar{V}^l_{\dot\alpha\dot\beta}
=\frac12({\tt V}^{-k}{\tt V}^{-l})\,.
\lb{A11}
\eea

The one-to-one correspondence between the specific variables used in
\cite{BN,CKR,CKO} and our variables in the tensor notation is
as follows
\footnote{Sometimes, for brevity, we omit the
antisymmetric tensor indices. }:
\bea
&&T^k={\tt F}^k-i{\tt G}^k\,,\quad
\tilde{T}^k=\tilde{\tt F}^k-i\tilde{\tt G}^k \,, \nn
&&
T^{*k}={\tt F}^k+i{\tt G}^k\,,\quad\widetilde{T^{*k}}=\tilde{\tt F}^k
+i\tilde{\tt G}^k\,,\\
&&\delta_\omega T^k=i\omega^{kl} T^l\,,\;\delta_\omega \tilde{T}^k=
i\omega^{kl}
\tilde{T}^l\,,\;\delta_\omega T^{*k}=-i\omega^{kl} T^{*l}\,,\;
\delta_\omega \widetilde{T^{*k}}=-i\omega^{kl} \widetilde{T^{*l}}\,,
\eea
where $\omega^{kl}=\xi^{kl}+i\eta^{kl}$ are the $U(N)$ parameters.
The relations between the self-dual (and anti-self-dual) parts of these
two sets of complex variables can be listed as
\bea
&&T^{+k}=\frac12(T+i\tilde{T})^k=({\tt F}-{\tt V})^k+i(\tilde{\tt F}
-\tilde{\tt V})^k
=2({\tt F}^{+k}-{\tt V}^{+k})\,,\lb{A15}\\
&&
T^{-k}=\frac12(T-i\tilde{T})^k=2V^{-k}\,,\lb{A16}\\
&&T^{*+k}\equiv\bar{T}^{+k}=(T^-)^{*k} = \frac12(T^*+i\widetilde{T^*})^k
=2{\tt V}^{+k}\,,\lb{A17}\\
&&T^{*-k}\equiv\bar{T}^{-k}=(T^+)^{*k}=\frac12(T^*-i\widetilde{T^*})^k
=2({\tt F}^{-k}-{\tt V}^{-k})\,.\lb{A18}
\eea

Being cast in the tensor notations, our Lagrangian \p{LFVN} becomes:
\bea
{\cal L}({\tt F},{\tt V})&=&-\frac14[({\tt F}^{+k}{\tt F}^{+k})+({\tt F}^{-k}
{\tt F}^{-k})]+
\frac12[({\tt V}^{+k}-{\tt F}^{+k})^2+({\tt V}^{-k}-{\tt F}^{-k})^2]\nn
&& +\,{\cal E}(A_n)\,,\lb{TFVN}
\eea
where the invariant interaction ${\cal E}$ is defined in terms of the
matrix
\be
a^{kl}=\frac14({\tt V}^{-k}{\tt V}^{-r})({\tt V}^{+r}{\tt V}^{+l})\,,\quad
A_n=\frac1n\mbox{Tr}a^n\,.
\ee

The tensor form of our algebraic equation of motion \p{EFVN} reads
\bea
({\tt F}^+-{\tt V}^+)^k_{mn}=\frac{\partial{\cal E}}{\partial
{\tt V}^{+kmn}}\,.\lb{ours}
\eea
After passing to the $T$-tensor notation according to Eqs. \p{A16}, \p{A17}, we can rewrite the
same equation as
\bea
T^{+k}_{mn}=4\frac{\partial{\cal E}}{\partial \bar{T}^{+kmn}}\,,\lb{tensalg}
\eea
that precisely coincides with the general twisted self-duality
condition postulated in \cite{BN,CKR}.
Our interaction function ${\cal E}$ proves to be
identical to the  ``deformation function'' ${\cal I}^{(1)}$ of this approach.
Note that the variables $T^k,\quad T^{*k}$ are entirely equivalent
to our variables  on mass shell, when the auxiliary fields are traded for the Maxwell field strengths by their equations of motion,
while off shell  it is more convenient to deal with the variables ${\tt F}^k, {\tt V}^k\,$.

 

\begin{thebibliography}{99}


\bibitem{GZ}M.K. Gaillard and B. Zumino, {\it Duality rotations for interacting
fields}, Nucl. Phys. B {\bf 193} (1981) 221.
\bibitem{GZ2}M.K. Gaillard and B. Zumino, {\it Self-duality in nonlinear
electromagnetism}, In: Supersymmetry and quantum field theory, eds.
J. Wess and V.P. Akulov, p. 121-129,
Springer-Vellag, 1998, {\tt hep-th/9705226};\\
M.K. Gaillard and B. Zumino, {\it Nonlinear electromagnetic
self-duality and Legendre transform}, In: Duality and Supersymmetric
Theories, eds. D.I. Olive and P.C. West, p. 33, Cambridge University
Press, 1999, {\tt hep-th/9712103}.
\bibitem{KT} S.M. Kuzenko and S. Theisen, {\it Supersymmetric duality
rotation}, JHEP {\bf 0003} (2000) 034, {\tt hep-th/0001068};\\
S.M. Kuzenko and S. Theisen, {\it Nonlinear self-duality and
supersymmetry}, Fortsch. Phys. {\bf 49} (2001) 273, {\tt
hep-th/0007231}.
\bibitem{ABMZ}
P. Aschieri, D. Brace, B. Morariu, B. Zumino,
{\it Nonlinear self-duality in even dimentions}, Nucl. Phys. B {\bf 574} (2000)
571, {\tt hep-th/9909021};\\
P. Aschieri, D. Brace, B. Morariu, B. Zumino,
{\it Proof of a symmetrized trace conjecture for the abelian
Born-Infeld Lagrangian}, Nucl. Phys. B {\bf 588} (2000) 521, {\tt
hep-th/0003228}.
\bibitem{AFZ}P. Aschieri, S. Ferrara and B. Zumino, {\it Duality
rotations in nonlinear electrodynamics and extended supergravity},
Riv. Nuovo Cim. {\bf 31} (2008) 625, {\tt arXiv:0807.4039 [hep-th]}.
\bibitem{BHN}G. Bossard, C. Hillmann, H. Nicolai, {\it $E_{7(7)}$ symmetry in perturbatively
quantised ${\cal N}=8$ supergravity}, JHEP {\bf 1012} (2010) 052,
{\tt arXiv:1007.5472 [hep-th]}.
\bibitem{Ka} R. Kallosh, {\it $E_{7(7)}$ symmetry and finiteness of ${\cal N}=8$ supergravity},  JHEP {\bf 1203} (2012) 083, {\tt arXiv:1103.4115 [hep-th]};
R. Kallosh, {\it ${\cal N}=8$ counterterms and $E_{7(7)}$ current
conservation}, JHEP {\bf 1106} (2011) 073, {\tt arXiv:1104.5480
[hep-th]}.
\bibitem{BN} G. Bossard, H. Nicolai, {\it Counterterms vs. dualities},
JHEP {\bf 1108} (2011) 074, {\tt arXiv:1105.1273 [hep-th]}.
\bibitem{IZ1} E.A. Ivanov, B.M. Zupnik, {\it New representation for Lagrangians
 of self-dual nonlinear electrodynamics}, In: Supersymmetries and
quantum symmetries, eds. E. Ivanov {\it et al}, p. 235, Dubna, 2002, {\tt
hep-th/0202203}.
\bibitem{IZ} E.A. Ivanov, B.M. Zupnik, {\it ${\cal N}=3$ supersymmetric Born-Infeld
theory}, Nucl. Phys. B {\bf 618} (2001) 3, {\tt hep-th/0110074}.
\bibitem{IZ2}
 E.A. Ivanov, B.M. Zupnik, {\it New approach to nonlinear
electrodynamics: dualities as symmetries of interaction}, Yader.
Fiz. {\bf 67} (2004)  2212 [Phys.  Atom. Nucl.  {\bf 67} (2004)
2188], {\tt hep-th/0303192}.
\bibitem{IZ3}E.A. Ivanov, B.M. Zupnik, {\it Bispinor auxiliary fields in
 duality-invariant electrodynamics revisited}, Phys. Rev. D {\bf 87} (2013) 065023, {\tt arXiv:1212.6637 [hep-th]}.
\bibitem{Ga} F.R. Gantmakher, {\it Theory of matrices}, Moscow, Nauka, 1967
(in Russian).
\bibitem{ZO} B.M.Zupnik, V.I.Ogievetsky, {\it Investigation of nonlinear
realizations of chiral groups by generating-functions method}, Theor. Math. Phys.
 {\bf 1} (1969) 14.
\bibitem{CKR}J.J.M. Carrasco, R. Kallosh, R. Roiban, {\it Covariant procedure
for perturbative nonlinear deformation of duality-invariant theories},
Phys. Rev. D {\bf 85} (2012) 025007, {\tt arXiv:1108.4390 [hep-th]}.
\bibitem{CKO}W. Chemissany, R. Kallosh, T. Ortin, {\it Born-Infeld with higher derivatives},
Phys. Rev. D {\bf 85} (2012) 046002, {\tt arXiv:1112.0332 [hep-th]}.
\bibitem{ILZ} E. Ivanov, O. Lechtenfeld,  B. Zupnik, {\it Auxiliary superfields
 in  ${\cal N}=1$  supersymmetric self-dual electrodynamics}, JHEP {\bf 1305} (2013) 133, {\tt arXiv:1303.5962 [hep-th]}.
\bibitem{Ku}S.M. Kuzenko, {\it Duality rotations in supersymmetric nonlinear
 electrodynamics revisited},  JHEP {\bf 1303} (2013) 153, {\tt arXiv:1301.5194 [hep-th]}.
\bibitem{Eucl} G.W. Gibbons, K. Hashimoto, {\it Non-Linear Electrodynamics in Curved Backgrounds},
JHEP {\bf 0009} (2000) 013, {\tt hep-th/0007019}.
\bibitem{Berm} D. Berman, {\it SL(2,Z) duality of Born-Infeld theory from non-linear self-dual electrodynamics in 6 dimensions},
Phys. Lett. B {\bf 409} (1997) 153, {\tt hep-th/9706208}.
\end{thebibliography}
\end{document}